\documentclass[prb,a4paper,10pt,twocolumn,showpacs,floatfix,superscriptaddress,amsmath,amsfonts,amssymb,preprintnumbers,longbibliography,citeautoscript]{revtex4-1}

\usepackage[T1]{fontenc}
\usepackage[utf8]{inputenc}
\usepackage{dcolumn,graphicx,color,booktabs,microtype,afterpage}
\usepackage[charter,greekuppercase=italicized]{mathdesign}

\renewcommand{\figurename}{Fig.}
\renewcommand{\tablename}{Table}
\makeatletter\renewcommand{\fnum@figure}[1]{\figurename~\thefigure.\ }\makeatother
\makeatletter\renewcommand{\fnum@table}[1]{\tablename~\thetable.}\makeatother

\graphicspath{{./}{figure/}}

\newcount\hh \newcount\mm
\hh=\time \divide\hh by 60
\mm=\hh \multiply\mm by 60 \mm=-\mm
\advance\mm by \time
\def\now{\number\hh:\ifnum\mm<10{}0\fi\number\mm}
\newcommand{\tcr}{\textcolor{black}}

\begin{document}

\makeatletter\renewcommand{\ps@plain}{%
\def\@evenhead{\hfill\itshape\rightmark}%
\def\@oddhead{\itshape\leftmark\hfill}%
\renewcommand{\@evenfoot}{\hfill\small{--~\thepage~--}\hfill}%
\renewcommand{\@oddfoot}{\hfill\small{--~\thepage~--}\hfill}%
}\makeatother\pagestyle{plain}

\title{Pressure effects on the electronic properties of the undoped superconductor ThFeAsN}

\author{N.\ Barbero}
\email{nbarbero@phys.ethz.ch}
\affiliation{Laboratorium f\"ur Festk\"orperphysik, ETH Z\"urich, CH-8093 Zurich, Switzerland}

\author{S.\ Holenstein}
\affiliation{Paul Scherrer Institut, CH-5232 Villigen PSI, Switzerland}%
\affiliation{Physik-Institut der Universit\"at Z\"urich, Wintherturerstrasse 190, CH-8057, Z\"urich, Switzerland}%

\author{T.\ Shang}
\affiliation{Paul Scherrer Institut, CH-5232 Villigen PSI, Switzerland}
\affiliation{Institute of Condensed Matter Physics, EPFL Lausanne, CH-1015 Lausanne, Switzerland}

\author{Z.\ Shermadini}
\affiliation{Paul Scherrer Institut, CH-5232 Villigen PSI, Switzerland}%

\author{F. Lochner}
\affiliation{Max-Planck-Institut f\"ur Eisenforschung, D-40237 D\"usseldorf, Germany}
\affiliation{Theoretische Physik III, Ruhr-Universit\"at, D-44801 Bochum, Germany}

\author{I. Eremin}
\affiliation{Theoretische Physik III, Ruhr-Universit\"at, D-44801 Bochum, Germany}

\author{C.\ Wang}
\affiliation{Department of Physics, Shandong University of Technology, Zibo 255049, China}

\author{G.-H.\ Cao}
\affiliation{Department of Physics and State Key Lab of Silicon Materials, Zhejiang University, Hangzhou 310027, China}
\affiliation{Collaborative Innovation Center of Advanced Microstructures, Nanjing 210093, China}

\author{R.\ Khasanov}
\affiliation{Paul Scherrer Institut, CH-5232 Villigen PSI, Switzerland}%

\author{H.-R.\ Ott}
\affiliation{Laboratorium f\"ur Festk\"orperphysik, ETH Z\"urich, CH-8093 Zurich, Switzerland}
\affiliation{Paul Scherrer Institut, CH-5232 Villigen PSI, Switzerland}

\author{J.\ Mesot}
\affiliation{Laboratorium f\"ur Festk\"orperphysik, ETH Z\"urich, CH-8093 Zurich, Switzerland}
\affiliation{Paul Scherrer Institut, CH-5232 Villigen PSI, Switzerland}

\author{T.\ Shiroka}
\affiliation{Laboratorium f\"ur Festk\"orperphysik, ETH Z\"urich, CH-8093 Zurich, Switzerland}
\affiliation{Paul Scherrer Institut, CH-5232 Villigen PSI, Switzerland}

\begin{abstract}
\noindent
The recently synthesized ThFeAsN iron-pnictide superconductor exhibits a $T_c$ of 30\,K, 
the highest of the 1111-type series in absence of chemical doping. 
To understand how pressure affects its electronic properties, we carried out microscopic investigations up to 3\,GPa 
via magnetization, nuclear magnetic resonance, and muon-spin rotation experiments.
The temperature dependence of the ${}^{75}$As Knight shift, the spin-lattice 
relaxation rates, and the magnetic penetration depth suggest a multi-band 
$s^{\pm}$-wave gap symmetry in the dirty limit, while the gap-to-$T_c$ ratio $\Delta/k_\mathrm{B}T_c$ hints 
at a strong-coupling scenario. Pressure modulates 
the geometrical parameters, thus reducing $T_c$, as well as $T_m$, 
the temperature where magnetic-relaxation rates are maximized, both at the same rate of approximately
--1.1\,K/GPa. This decrease of $T_c$ with pressure is consistent with band-structure calculations, 
which relate it to the deformation of the Fe 3$d_{z^2}$ orbitals.
\end{abstract}


\maketitle\enlargethispage{7pt}

\textit{Introduction.} 
The doping-induced superconductivity below $T_c$ = 26\,K in LaFeAsO$_{1-x}$F$_x$\cite{Kamihara2008} 
triggered long-term research interests towards iron-based superconductors (FeSC), further boosted 
by the $T_c = 55$\,K of SmFeAsO$_{1-\delta}$\cite{Ren2008}. Recently, we reported on 
superconducting properties of ThFeAsN,\cite{Shiroka2017}  an undoped FeSC with a remarkable 
$T_c$ of 30\,K. Our data indicate that Fermi-surface modifications due to structural distortions and 
correlation effects may be as effective as doping in suppressing the antiferromagnetic order 
in favor of the formation of a superconducting phase. %
This is in contrast with %
most other REFeAsO-type compounds (RE = rare earth), where the %
quaternary parent compounds usually order magnetically and superconductivity 
is %
established via F-\cite{Wen2008, Sadovskii2008, Paglione2010, Johnston2010, Si2016} 
or H-doping.\cite{Muraba2014,Muraba2015}
Due to strong electron correlations (compared with kinetic energy), 
iron pnictides are intermediately coupled systems. For this reason, the experimental
values of $T_c$ are distinctly higher than those calculated by assuming the 
electron-phonon coupling mechanism\cite{Boeri2008}, which claims $T_c$ values below 1\,K. 
Among the strong-correlation effects, antiferromagnetic (AFM) spin fluctuations are widely accepted to mediate the SC pairing but the detailed interaction model and an unequivocal identification of the gap symmetry are still being debated.\cite{Bang2017}

In an attempt to establish \textit{(i)} what causes the suppression of AFM
order in nominally undoped FeSC compounds, \textit{(ii)} why they become superconductors, 
and \textit{(iii)} what determines their $T_{c}$ values, we investigated ThFeAsN under 
applied hydrostatic pressure using different local probes.
Hydrostatic and/or chemical pressure modify the structure 
and thus tune the $T_c$ of iron-based superconductors, such as FeSe, whose 
original $T_c = 8.5$\,K increases to 36.7\,K at 8.9\,GPa.\cite{Medvedev2009}
In particular, hydrostatic pressure is regarded as a \textit{clean} tuning parameter for 
studying the effects of structural distortions on the electronic properties. %
A dependence of $T_c$ on the crystallographic As-Fe-As bond angle\cite{Lee2008} or on the anion  %
height above the iron layers\cite{Mizuguchi2010} $h_\mathrm{Pn}$ (see Fig.~1 in Mizuguchi 
et al.\cite{Mizuguchi2010}) has previously been noted. 
With $a = 4.037$\,\AA\ and $c =  8.526$\,\AA,\cite{Wang2016} the tetragonal 
($P4/nmm$) structure of ThFeAsN %
implies an $h_\mathrm{Pn}$ = 1.305(4)\,\AA, lower than the 
optimum anion height $h_\mathrm{Pn}^\mathrm{opt} = 1.38$\,\AA.\cite{Mizuguchi2010}
Hence, in the case of ThFeAsN, structural deformations induced by hydrostatic pressure would invariably 
lower $T_c$,\cite{Wang2016} in contrast to the above mentioned FeSe case. %
To test this hypothesis and understand how pressure affects the electronic properties 
of an undoped 1111 superconducting compound, we performed magnetization-, nuclear 
magnetic resonance (NMR) and muon-spin rotation ($\mu$SR) measurements on 
ThFeAsN under applied pressures up to 3\,GPa.

First, we confirm experimentally the expected reduction of $T_c$ with pressure. Then, on account of the $T$-dependence of 
the NMR Knight-shifts and spin-lattice relaxation rates, as well as $\mu$SR relaxation rates, we argue that the energy gap $\Delta$ of superconducting ThFeAsN adopts the $s^{\pm}$ symmetry, which %
persists up to at least 1.47\,GPa. In the same pressure region, the ratio $R = \Delta/k_\mathrm{B}T_c$ is reduced continuously from 2.16(3) at ambient pressure to 1.82(3) at 2.48(2)\,GPa, thus exceeding the BCS weak-coupling value of 1.76.
The moderate variation of $T_c$ with pressure is corroborated by results of band-structure calculations which imply only tiny changes in the electronic excitation spectrum around $E_\mathrm{F}$. %
The abrupt quenching of magnetic excitations, as indicated by a cusp in $1/T_1T(T)$ at $T_m$ > $T_c$, persists upon increasing pressure and $T_m$ is reduced at the same rate as $T_c$.

\textit{Synthesis and preliminary characterization.} The polycrystalline ThFeAsN sample was synthesized via high-temperature 
solid-state reaction, as reported in Ref.~\onlinecite{Wang2016}.
X-ray diffraction and energy-dispersive x-ray measurements confirmed the absence of 
spurious phases (within $\sim$1\%).

\begin{figure}[th]
\includegraphics[width=0.47\textwidth]{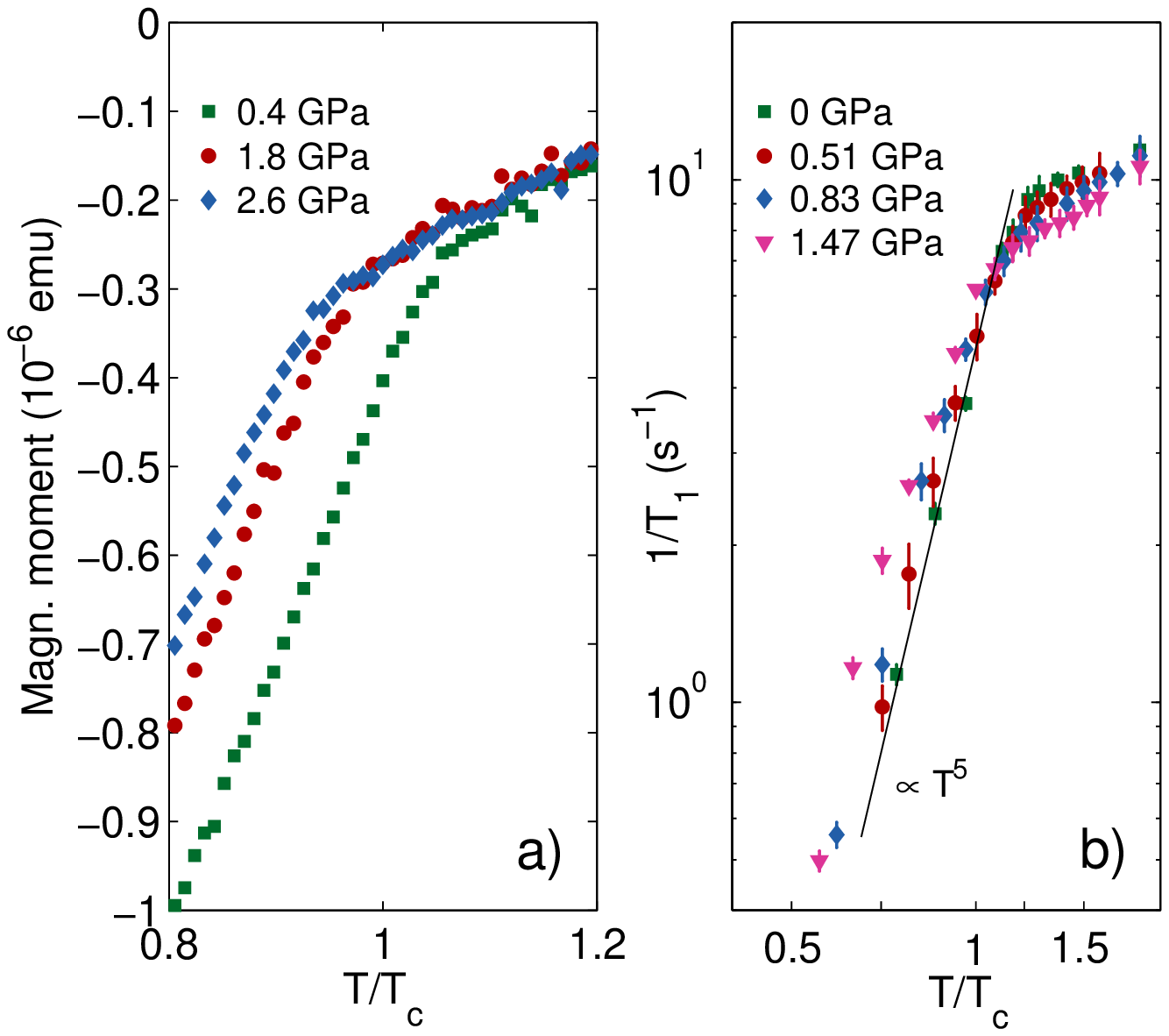}
\includegraphics[width=0.48\textwidth]{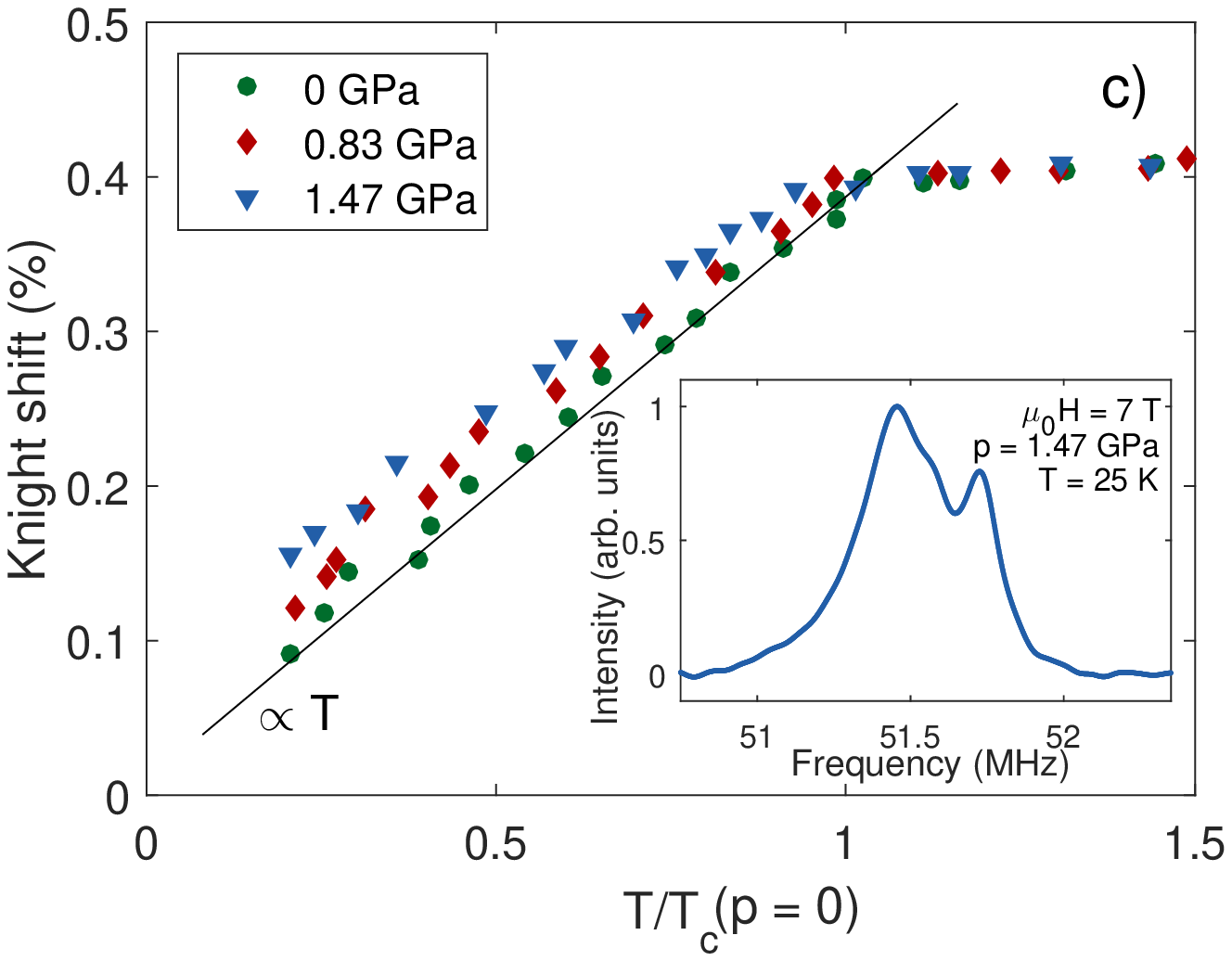}
\caption{\label{fig:fig01}(a) Temperature-dependence of magnetization for 
selected applied pressures, measured at $\mu_{0}H = 2$\,mT. 
(b) ${}^{75}$As NMR $1/T_1(T)$ 
data at ambient- and at selected hydrostatic pressures. The line represents a $T^{5}$ behavior of relaxation, with the exponent decreasing
down to 3.6 at 1.47\,GPa. To improve the readability of the plot, we are not indicating the location of $T_m$ and $T_c$ values, but we report them in Fig.~\ref{fig:fig02}.
(c) Temperature dependence of $^{75}$As NMR Knight-shift 
at three selected pressures. Uncertainties are of the order of the marker size. 
Inset: the $^{75}$As NMR signal measured at 1.47\,GPa, 7.06\,T, and 25\,K.}
\end{figure}

\vspace{5pt}
\textit{Magnetization measurements under applied pressure.} The magnetization measurements 
were performed with a superconducting quantum interference device (SQUID) %
MPMS XL magnetometer. Preliminary measurements at ambient pressure 
revealed the presence of a tiny quantity of impurities ($\sim$ 0.18\%, assuming that they are
of ferromagnetic nature).\cite{Wang2016, Shiroka2017} This, along with a broad drop-down in $M(T)$ data 
 below $T_c$, related to defect-induced disorder, 
suggest that ThFeAsN in the SC phase should be described by models in the 
dirty limit. 
Hydrostatic pressures up to 3.1\,GPa were achieved by means of a home-made 
diamond-anvil cell with a beryllium-copper (BeCu) body.
We chose Daphne Oil 7575 as the pressure-transmitting medium and a piece of 
lead to monitor the pressure.\cite{Eiling1981}
For the magnetometry measurements we used a tiny piece of ThFeAsN 
($m \sim 40$\,$\mu$g), whose magnetic response was of the order of 
1.5--3\,$\mu$emu. Because of the tiny signal, each measurement was 
performed with a background-subtraction procedure. 
Typical magnetization data at different applied pressures are shown in 
Fig.~\ref{fig:fig01}(a). The linearly decreasing trend of $T_{c}$, as 
\begin{figure}[th]
\includegraphics[width=0.47\textwidth]{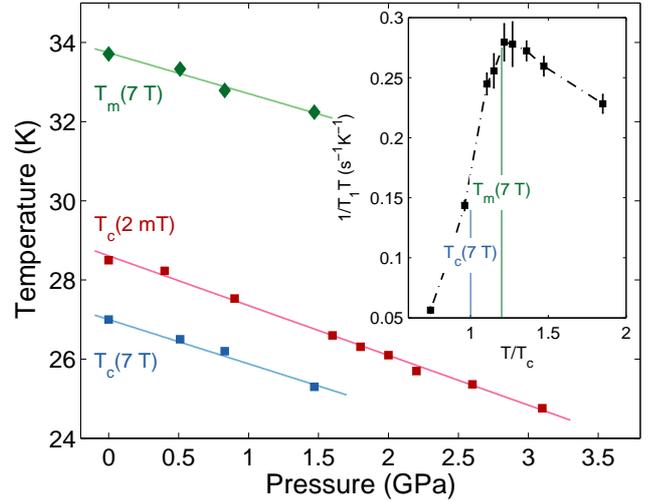}
\caption{\label{fig:fig02}Pressure dependence of $T_m$ and $T_c$
as determined via magnetometry and NMR measurements. 
$T_m$ (green diamonds) refers to the temperature where the electronic relaxation rates
are maximized. The inset shows the temperature dependence of $1/T_1T$ at ambient pressure, highlighting
the maximum at $T_m$ and a kink at $T_c$. Recently, a similar but broader feature in $1/T_1T$ of FeSe was attributed
to a pseudogap behavior.\cite{Shi2018} Note that for ThFeAsN $T_m \sim 1.2T_c$, at least up to 1.47\,GPa.}
\end{figure}
determined from magnetization data, is shown in Fig.~\ref{fig:fig02} 
(red squares) and 
agrees well with the prediction of a reduced $T_c$ %
at lower anion heights. A linear fit within the explored pressure range gives 
a slope of $\partial T_c/\partial p = -1.12 \pm 0.02$\,K/GPa, similar 
to $-1.5$\,K/GPa found in LiFeAs,\cite{Gooch2009} another %
iron-based superconductor without doping.

\textit{NMR measurements under applied pressure.}  NMR measurements up to 
1.47\,GPa were performed using a BeCu 
piston-clamped high-pressure cell. The $^{75}$As NMR investigations included line- and spin-lattice 
relaxation time ($T_1$) measurements in a magnetic field of 7.06\,T.\footnote{$^{75}$As, 
with spin $I$ = 3/2 and Larmor frequency $\nu_\mathrm{L}$ = 51.523\,MHz at 
7.06\,T, was chosen because occupies a single site and is sensitive to
 the structural- and electronic variation in the FeAs layers under pressure.
The $^{75}$As NMR spectra were obtained via fast Fourier transformation of 
spin-echo signals 
generated by $\pi/2$--$\pi$ rf pulses of 2 and 4\,$\mu$s, respectively, 
with recycle delays ranging from 0.1\,s at room temperature up to 6\,s 
at 5\,K and echo times of 50\,$\mu$s. Given the long rf-pulse 
length, frequency sweeps in 40-kHz steps were used to cover the 
spectrum central transition ($\sim 1$\,MHz wide).}
$T_1$ values measured at both peaks 
of the central-transition line via inversion recovery resulted identical. Pressure was monitored 
\textit{in situ} by using the nuclear quadrupolar resonance signal of ${}^{63}$Cu 
in Cu$_2$O.\cite{Reyes1992}

A typical $^{75}$As NMR line at 7.06\,T is shown in the inset of 
Fig.~\ref{fig:fig01}(c). Due to the large quadrupole moment of $^{75}$As  
($Q = 31.4$\,fm$^2$), we considered only the central component of the NMR spectrum, 
which exhibits a typical second-order powder pattern with dipolar broadening.
For temperatures from 4 to 295\,K and hydrostatic pressures from 
zero up to 1.47\,GPa, the central-line transition \tcr{exhibits minor changes} in shape and position.
The spectra were fitted using the quadrupolar exact software 
(QUEST)\cite{Perras2012}, assuming no planar anisotropy 
($\eta = 0$, as from experimental observations) and obtaining typical quadrupolar 
frequencies $\nu_Q$ of $\sim$5.6\,MHz. The full width at half maximum (not shown) is negligibly affected by temperature or pressure, thus confirming the absence\cite{Shiroka2017,Albedah2017} of AFM long-range order, which would 
otherwise result in a remarkable broadening of the spectral lines starting at the onset of the transition.

Figure~\ref{fig:fig01}(c) shows the Knight shift $K_s(T) = (\nu - \nu_\mathrm{L})/\nu_\mathrm{L}$ values as a function of temperature. At all the applied pressures, $K_s(T)$ exhibits 
a linearly decreasing trend below $T_{c}$, compatible with an $s^{\pm}$-wave 
scenario.\cite{Bang2017} In fact, $K_s(T) \sim \mathrm{Re} \chi_S(q=0,\omega \rightarrow 0)$,  
i.e., in the uniform susceptibility limit ($q=0$), the inter-band 
scattering is suppressed and the Knight-shift value %
includes %
only the independent contributions from the hole- and electron bands.\cite{Bang2008, Bang2017} 
In the clean limit, this implies an exponential temperature-dependence 
for $K_s(T)$ in the $s^{\pm}$-wave case. However, as confirmed by magnetization 
data, our sample is not free of impurities. As reported in the literature,\cite{Bang2008, Kawabata2008, Matano2008, Nakai2010} 
impurity self-energies form resonance states inside the SC gap and, thereby, affect the functional form of $K_s(T)$. The results of these calculations are compatible with the 
linear trend we observe. From the Knight-shift perspective, a 
dirty $s^{\pm}$-wave superconductor exhibits the features of a clean $d$-wave SC, but the latter interpretation is 
ruled out by the $\mu$SR data (see below). 
Quantitatively, the impurity effect is measured in terms of the 
$r \equiv \Gamma/\Delta$ ratio, where $\Gamma$ is the impurity scattering 
rate and $\Delta$ the SC gap value. The critical value separating the exponential 
from linear behavior is $r_\mathrm{cr} = 0.045$.\cite{Bang2008, Bang2017} 
Although we cannot provide an estimate for the $r$-value, the observed 
linear dependence of $K_\mathrm{s}(T)$ suggests that, in our case, $r > r_\mathrm{cr}$.

From the peak separation of the second-order quad\-ru\-pole\--broad\-en\-ed powder 
spectra  of the central transition [see Fig.~\ref{fig:fig01}(c) inset], 
$\Delta f = f_\mathrm{right} - f_\mathrm{left}$, one can evaluate the 
electric-field gradient component $eq = V_{zz} = {2I(2I-1)\,h\nu_Q}/{(3eQ)}$. 
Here $\nu_Q$ is obtained from the quad\-ru\-pole splitting frequency via %
\begin{equation} \label{eq:diffreq}
\Delta f = \frac{\frac{25}{9}\left[I(I+1)-\frac{3}{4}\right]\nu_Q^2}{16 \nu_L}.
\end{equation}
We observe that pressure reduces the distance between the peaks. 
This implies a slight \textit{symmetry enhancement} 
upon increasing pressure, resulting in $eq \equiv V_{zz}$ values of 
$-1.65 \times 10^{21}$ and $-1.45 \times 10^{21}$\,V/m$^2$ at ambient 
pressure and 1.47\,GPa, respectively.

To study the electron-spin dynamics, the temperature dependence of the 
nuclear spin-lattice relaxation rate $1/T_1$ was measured at different 
pressures, as shown in Fig.~\ref{fig:fig01}(b). 
At all pressures, $1/T_1(T)$ exhibits a kink, resulting in a maximum at  $T_m$, 
if plotted as $1/(T_1T)$ (see inset of Fig.~\ref{fig:fig02}). The data above $T_m$ 
reveal an additional relaxation channel due to short-range AFM spin fluctuations, 
as shown by M\"ossbauer\cite{Albedah2017} and ambient-pressure NMR\cite{Shiroka2017} 
results, the latter extending to values of $T/T_m < 0.5$. Below $T_m$ this relaxation 
channel is increasingly inhibited before the onset of superconductivity at $T_c$, which 
reduces $1/(T_1T)$ even further. Note that, as shown in Fig.~\ref{fig:fig02}, 
$T_m(p)$ decreases monotonously by $-1.0 \pm 0.1$\,K/GPa, i.e., virtually with 
the same slope as ${\partial T_{c}}/{\partial p}$. The $T_c$ values (onset of SC transition) were determined from 
2-mT magnetization measurements (red squares) and from the maxima of the $T$-derivative 
of the NMR spin-lattice relaxation data 
at 7.06\,T (blue squares), as described in the Supplemental Material.\footnote{See Supplemental Material at [URL will be inserted by publisher] for the plot of the $T$-derivative of the $1/T_1T(T)$ curve and the definitions of $T_m$ and $T_c$.} 

In several 1111 FeSCs, the $T^3$-dependence\cite{Kawasaki2008,Mukuda2009} of $1/T_1(T)$ suggests a nodal gap,
however, ruled out by angle-resolved photo-emission 
spectroscopy experiments.\cite{Ding2008} By employing $\mathcal{T}$-matrix 
theory, the impurity-scattering effect has been included in the modeling 
of the $1/T_1(T)$ dependence.\cite{Ziegler1996} %
For $r = 0$, one obtains the exponential behavior of $1/T_1(T)$ expected for an 
$s^{\pm}$-wave,\cite{Oka2012} whereas for $r = r_\mathrm{cr}$ one finds 
$1/T_1(T) \propto T^3$ and a suppression of the 
Hebel-Slichter coherence peak,\cite{Bang2009} in good agreement with experiments.\cite{Kawasaki2008} 
Our data, shown in Fig.~\ref{fig:fig01}(b), 
exhibit an even steeper $T$-dependence of the type $1/T_1(T) \propto T^5$, 
as previously reported for other FeSCs.\cite{Nakai2010, Matano2009} 
Such a $T^5$-behavior does not require a different gap symmetry, if strong-coupling effects 
are taken into account, as discussed in the introduction. According to theoretical estimates, a $R \equiv \Delta/k_\mathrm{B}T_c = 2.5$ ratio in a dirty-limit sample with $r = r_\mathrm{cr}$ gives rise to the observed $T^5$ power-law 
behavior.\cite{Bang2009} The reduction of the power-law exponent from 5 to 3.6, for pressures 
from zero to 1.47\,GPa, indicates a pressure-induced weakening of the coupling, in good agreement with our $\mu$SR results (see below). The same
results also rule out a clean $d$-wave superconducitivity scenario and are compatible with an iso-/anisotropic $s$- or 
$s^{\pm}$-wave model for ThFeAsN.\cite{Shiroka2017,Adroja2017} 

\begin{figure}[t]
\includegraphics[width=0.46\textwidth]{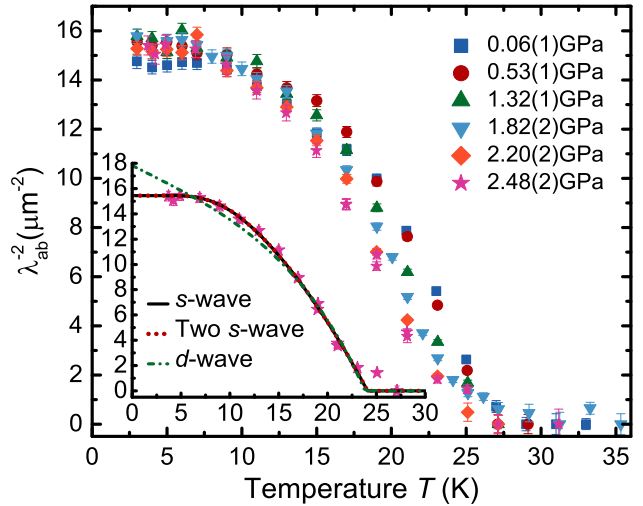}
\caption{\label{fig:fig03}Temperature dependence of $\lambda_{ab}^{-2}$ as 
extracted from TF-$\mu$SR measurements show a suppression of $T_{c}$ 
with pressure, but no changes in functional form. Inset: Fits of 
$\lambda_{ab}^{-2}(T)$ data taken at the highest pressure by using %
different SC models. Note that the one-gap and the two-gap $s$-wave models overlap perfectly.}
\end{figure}

\vspace{7mm}
\textit{Transverse-field (TF) $\mu$SR measurements under high pressure.} The $\mu$SR investigations 
were performed at the GPS (ambient pressure) and the GPD (high-pressure) spectrometers 
of the Paul Scherrer Institute, Villigen. 
Since the high-pressure measurements require a relatively large sample 
mass ($\sim$ 2\,g), a new polycrystalline sample with $T_c = 27$\,K 
was prepared. The 
lower $T_{c}$ is due to a different preparation protocol.
The GPS measurements on the new batch confirmed the earlier 
findings\cite{Shiroka2017} and were used as reference to analyze 
the high-pressure data. 
The muon fraction stopping in the pressure cell ($f_\mathrm{cell} = 60$\%) 
was determined by fitting a zero-field (ZF) spectrum with the cell 
relaxation rates fixed to their literature values\cite{Khasanov2016} 
and the sample relaxation rate fixed to the GPS value, hence leaving the 
muon stopping fraction as the only free parameter. 
The absence of significant changes with temperature in the ZF relaxation 
rate of the sample, even at the highest pressure, rule out a possible 
pressure-induced magnetic order.
Thus, we focused on the TF measurements in the SC region, carried out at 70\,mT. 
The data were analyzed using:
\begin{align*}
\label{eq:fit_background}
A(t)/A_0 &= (1-f_\mathrm{cell})\cos(\gamma_{\mu}B_\mathrm{sc}t + \phi)\exp(-\lambda_\mathrm{sc}t-\sigma_\mathrm{sc}^2t^2/2) \\ \nonumber
&+ f_\mathrm{cell}\cos(\gamma_{\mu}B_\mathrm{cell}t + \phi)\exp(-\lambda_\mathrm{cell}t-\sigma_\mathrm{cell}^2t^2/2),
\end{align*}
where $A_0$ is the initial asymmetry, $\gamma_{\mu}$ the muon gyromagnetic ratio, $B$ the local field at the muon stopping site, $\phi$ the initial phase, and $\lambda$ and $\sigma$ the exponential and Gaussian relaxation rates, whose subscript labels denote the parameters for muons stopping in the sample and the cell, respectively.
To ensure a robust fit, the change of $B_\mathrm{cell}$ and $\sigma_\mathrm{cell}$ was related to the field shift in the sample relative to 
$B_\mathrm{ext}$:\cite{Maisuradze2011}
\begin{equation}\label{eq:fit_B}
\begin{split}
B_\mathrm{cell}(T) = B_\mathrm{ext} + c_1\left[B_\mathrm{ext} - B_\mathrm{sc}(T)\right], \\
\sigma_\mathrm{cell}^{2}(T) = \sigma_\mathrm{cell}^2(T > T_c) + c_2^2(B_\mathrm{ext}-B_\mathrm{sc})^2.
\end{split}
\end{equation}
where $c_1$ and $c_2$ are proportionality constants. Since $\lambda_\mathrm{cell}$ 
varies with temperature, %
its intrinsic $T$-dependence was determined by requiring that the zero-pressure GPD measurements reproduce the GPS results, from which we evaluated an average penetration depth $\lambda_{ab}$(0\,K) = 255(1)\,nm. As can be seen from the temperature dependence of the inverse-squared 
penetration depth,\footnote{the penetration depth was calculated from $\sigma_\mathrm{sc}(T)$ and is proportional 
to the superfluid density} hydrostatic pressure reduces the superconducting 
$T_c$ [at a rate of 1.1(2)\,K/GPa], while barely influencing the gap symmetry. 
The latter is clearly not of a clean $d$-wave type, as shown by the poor fit of 
the highest-pressure dataset (see inset of Fig.~\ref{fig:fig03}), while a dirty d-wave scenario is already excluded by the NMR measurements. Our
data fit with the same accuracy a single- %
or a two-gap $s$-wave model (the latter yields two gaps of almost equal magnitude). 
Hence, $\mu$SR data are compatible with an $s$-wave model, %
but there is no strong evidence for claiming ThFeAsN to be a double-gap 
superconductor. Interestingly, the gap value is suppressed 
faster [from 5.0(1)\,meV at 0.06(1)\,GPa to 3.8(1)\,meV at 2.48(2)\,GPa] 
than $T_c$, hence implying a reduction of the $\Delta(0)/k_{\mathrm{B}}T_c$ 
ratio from 2.16(3) to 1.82(3) upon increasing pressure, the latter being 
closer to the BCS value of 1.76.
The absence of a magnetic order even at the highest pressures suggests 
that the reduction of $T_c$ is of structural origin. The lack of AFM order, as confirmed by both NMR and 
zero-field $\mu$SR data, has no clear explanation.\cite{Shiroka2017} An educated guess\cite{Note3} 
suggests intrinsic disorder as the key reason to prevent AFM order. %

\begin{figure}[t]
\includegraphics[width=0.38\textwidth]{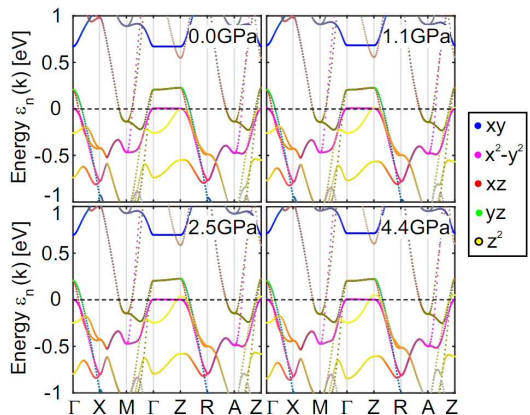}
\caption{\label{fig:fig04}ThFeAsN band-structure calculations upon increasing 
pressure. Note the lack of major changes at the Fermi-energy level, except for 
a minor change in the 3$d_{z^2}$ orbitals close to the $Z$ point.}
\end{figure}

\textit{Band-structure calculations.} For the DFT calculations we resorted to %
the Vienna \textit{ab initio} 
simulation package (VASP)\cite{Kresse1993,Kresse1996a,Kresse1996b} with the 
projector augmented wave (PAW)\cite{Bloechl1994} basis and made use of %
the measured ThFeAsN crystal parameters.\cite{Wang2016} 
Hereby, we use VASP in the generalized-gradient approximation (GGA)\cite{Perdew1996} 
and show in Fig.~\ref{fig:fig04} the resulting band-structure evolution 
with pressure. 
One can rule out substantial %
pressure effects on %
the electronic structure --- within the accuracy of band-structure 
calculations --- including here changes in the effective mass. 
For the experimentally accessible pressure  values, we find %
insignificant modifications of the electronic structure, except for a minor 
change in the 3$d_{z^2}$ orbitals close to the $Z$ point.

This lack of substantial pressure effects raises the question of 
the origin of $T_c$ suppression in this compound.  
We recall that the electronic properties of ThFeAsN are similar to those 
of LaFeAsO$_{0.9}$F$_{0.1}$ and that, as shown above, ThFeAsN
can be regarded as a superconductor in the dirty 
limit.
\footnote{We recall that the transport and magnetic properties 
of ThFeAsN are similar to those of LaFeAsO$_{0.9}$F$_{0.1}$\cite{Shiroka2017}, 
which indicates that the absence of a long-range magnetic order in the 
nominally undoped ThFeAsN can be due to intrinsic disorder.
It is known that the long-range antiferromagnetic order in 
the iron-based superconductors can be even destroyed by nonmagnetic 
impurities, despite the electron-band structure remaining unchanged with respect 
to the undoped case.\cite{Brouet2010,Dhaka2010}
The resulting phase diagram is similar to that obtained by introducing 
extra holes or electrons in the FeAs layers. Theoretically, this can be 
understood as a result of the stronger effect of nonmagnetic impurities 
on the AFM order than on the multiband $s^{\pm}$-wave superconductivity.\cite{Vavilov2011}
While both the intra- and the interband impurity 
scattering are destructive for the long-range AFM order, only the 
interband scattering is pair-breaking for an $s^{\pm}$ 
superconducting state.} %
In this case,  despite a nested electronic structure, the AFM ordering 
of excitonic type is more sensitive to disorder than is the $s^{\pm}$-wave 
superconductivity.\cite{Brouet2010,Dhaka2010, Vavilov2011, Lang2016}
Hence, in the nominally undoped compound, disorder tends to promote the 
$s^{\pm}$ states over the AFM ordering. Indeed, disorder affects the magnetic 
order by lowering $T_{\mathrm{N}}$, whereas, in case of superconductivity,  
the inter-/intraband scattering plays a major role and only the interband
impurity scattering lowers the $T_c$. This may explain why, in contrast to the 
cleaner LaFeAsO, ThFeAsN is a superconductor instead of an antiferromagnet, 
although both are nominally undoped compounds.

As reported in Ref.\ \onlinecite{Prando2015}, in doped LaFeAsO 
hydrostatic pressure does not influence $T_c$, although the superfluid density is enhanced. 
In fact, pressure seems to slightly change the ratio of intra- to inter\-band 
impurity scattering, without sensibly affecting $T_c$. The most plausible 
reason for the lowering of $T_{c}$ with pressure %
in ThFeAsN could be a subtle modification of the electronic structure (beyond 
density functional theory), which can account for the simultaneous 
suppression of the AFM fluctuations and of SC. 
For example, smaller Fermi-surface pockets (with respect to 
simulations) would imply a much stronger effect of pressure on the 
electronic structure. Nearly all 
iron-based superconductors, including the LaFeAsO family, exhibit an 
unusual renormalization of the electronic structure, which results in much 
smaller Fermi surface pockets than anticipated from DFT calculations.
\cite{Charnukha2015} By assuming such a renormalization, i.e., the 
so-called red/blue shift for the pockets, one may justify the suppression 
of AFM fluctuations and SC in ThFeAsN as well.

\vspace{3mm}
\textit{Conclusion.} %
In ThFeAsN, the Knight shift $K(T)$, the spin-lattice relaxation times $T_1(T)$, 
and the London penetration depth $\lambda(T)$ indicate that pressure reduces $T_c$ 
[$\partial T_c/\partial p = -1.12(2)$\,K/GPa] 
and weakens the pairing interaction, as measured by the ratio $\Delta/k_\mathrm{B}T_c$. 
Interestingly, $T_m(p)$ too is reduced by pressure at the same rate of $T_c$, confirming that
magnetic excitations which reflect AFM spin fluctuations, while competing with superconductivity, play an essential role in the pairing process. 
Finally, our experimental data and DFT calculations indicate an $s^{\pm}$ SC order parameter 
independent of pressure and suggest that intrinsic disorder plays a key role in suppressing 
antiferromagnetism in ThFeAsN.

\begin{acknowledgments}
%
The authors thank H.\ Luetkens for useful discussions. 
this work was financially supported in part by the Schweizerische 
Nationalfonds zur F\"{o}rderung der Wissenschaftlichen Forschung (SNF) and by the SNF
fund 200021-159736.
\end{acknowledgments}

%

\newpage

\section{Supplemental Material} 

We include the accurate description of the methods used to determine the values of $T_c$ and $T_m$ from the magnetization- (see Fig. 1a) and the NMR data (see Fig. 1b). The resulting metadata are plotted as a function of pressure in Fig. 2 of the
manuscript.

$T_c$ was evaluated from the magnetization data at the onset value, resulting from the crossing of two lines, i.e. the linear fit of the neighboring regions below (0.8 < $T/T_c$ < 1) and above (1 < $T/T_c$ < 1.2) the critical temperature, respectively at each applied pressure.

In the case of the NMR data, we plotted (see Fig. 1 of the Supplemental Material) the $T$-derivative of $1/T_1T$ and identified $T_c$ as maximum [inflection point in 1/$T_1T(T)$] and $T_m$ as zero of the function [maximum in 1/$T_1T(T)$].

\begin{figure}[th]
\includegraphics[width=0.47\textwidth]{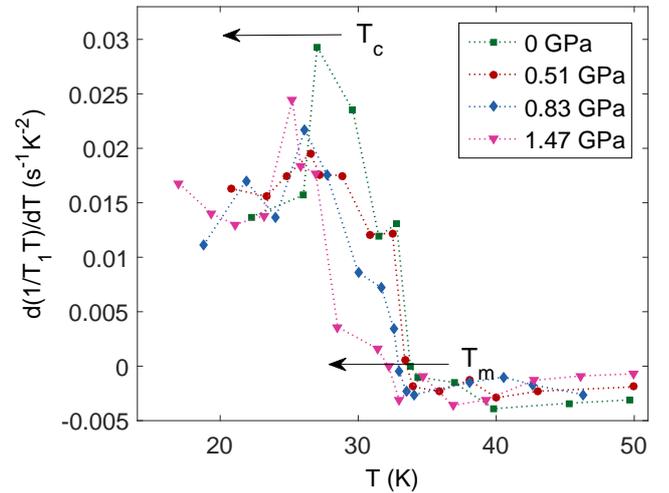}
\caption{\label{fig:fig01} From the ${}^{75}$As NMR $T_1$ values, we show the $T$-derivative of the $1/T_1T(T)$ curve for the four applied hydrostatic pressures. The decreasing trend for both $T_c$ and $T_m$ values upon increasing pressure is highlighted by two horizontal arrows. The dotted lines are a guide to the eye. Since the datasets are noisy, we empirically enhanced the accuracy in estimating the $T_c$ and $T_m$ values, by fitting these data with shape-preserving splines.}
\end{figure}

\end{document}